\documentclass[aps,pra,preprint]{revtex4}
\usepackage{graphicx}
\usepackage{latexsym}
\usepackage{amsmath}
\usepackage{amssymb}
\usepackage{float}
\usepackage{epstopdf}
\usepackage{bm}

\begin{document}
\baselineskip=22pt
\title{Analysis of Floquet formulation of time-dependent density-functional theory}
\author{Prasanjit Samal}
\email{Fax:+91-512-259 0914 ; E-mail: dilu@iitk.ac.in}
\author{Manoj K. Harbola}
\email{Fax:+91-512-259 0914 ; E-mail: mkh@iitk.ac.in}
\affiliation{Department of Physics, Indian Institute of Technology,
 Kanpur U.P. 208016, India}
\begin{abstract}
Floquet formulation of time-dependent density-functional theory is revisited in light
of its recent criticism [Maitra and Burke, Chem. Phys. Lett. 359 (2002), 237]. It is
shown that Floquet theory is well founded and its criticism has overlooked important
points of both the Runge-Gross formalism and Floquet formulation itself. We substantiate
our analysis by examples similar to that considered by Maitra and Burke. 
\end{abstract}
\maketitle
\newpage

Density functional theory (DFT) is a well established theory for the ground-states of 
Coulombic systems. It is conceptually simple and practically useful in various 
branches of physics, chemistry and material science. Based on the work of Hohenberg 
and Kohn (HK) \cite{hk} and Kohn and Sham (KS) \cite{ks}, DFT has become a practical 
tool for the electronic structure calculation of atoms, molecules and solids \cite{
parr2}.  Time-independent density functional formalism is also being extended 
to deal with excited-states \cite{ag1,ag2,ln,smss,gb,harb,shh,sh}. On the other hand,
time-dependent phenomena are not accessible through traditional DFT. Time-dependent 
density functional theory (TDDFT) \cite{dg,barto,rg} is the generalization of the ground-state 
DFT to time-dependent problems. The development of TDDFT is relatively recent. The central 
result of modern TDDFT is a set of time-dependent Kohn-Sham equations which are structurally 
similar to the time-dependent Hartree-Fock equations but include in principle exactly all 
the many-body effects through a local time-dependent exchange-correlation potential. TDDFT 
allows access to  lots of interesting and important phenomena which can't be easily dealt 
with in static DFT. It has become popular for studying atoms and molecules in intense laser
fields \cite{tong}, calculating the excitation spectra and oscillator strengths \cite{casida,
pgg}, frequency dependent polarizabilities, hyperpolarizabilities \cite{gisb} and 
optical response of solids, etc. \cite{primer}. TDDFT is especially very useful for quantum 
control problems \cite{rabitz}. This is because for intense laser fields where correlation 
effects are quite crucial, TDDFT appears to be the only way of studying the quantum 
dynamics of a many-electron system. \\

Initial steps towards the rigorous foundation of TDDFT were taken by Deb and Ghosh 
\cite{dg} and by Bartolotti \cite{barto}. They formulated and explored HK and KS 
type theorems for the time-dependent density. Each of these derivations, however, 
was restricted to a particular class of allowable time-dependent potentials (to 
potentials periodic in time). Because of the periodicity of the potential in time,
Deb and Ghosh developed TDDFT with the Floquet formulation. Further in recent times
Telnov and Chu \cite{chu1,chu2,chu3,chu4,chu5} have developed approaches for the
nonperturbative treatment of the strong field processes. These are the approaches
which involve the transformation of the TDDFT equations into an equivalent time-independent
generalized Floquet matrix eigenvalue problem. A variational perturbation method based 
on the Floquet theory has also been developed and applied \cite{arup} to study optical 
properties of atoms within TDDFT. This perturbative Floquet approach is suitable for the
treatment of weak-field processes. A general proof of density to potential mapping 
($\rho(\vec r,t)\to v(\vec r,t)$) for a time-dependent density was given by Runge and Gross 
\cite{rg}. Runge and Gross (RG) \cite{rg} formally established the TDDFT by showing that 
for a given initial state the evolving density $\rho(\vec r,t)$ uniquely 
determines the corresponding time-dependent potential $v(\vec r,t)$. Two features of RG 
formalism are noteworthy: one that the time-dependent wavefunction for a given density has 
functional dependence on both the density as well as the initial wavefunction $\Psi_0$ . 
Secondly the formalism is valid for potentials that are Taylor series expandable at the 
initial time $t = t_0$. This makes the formalism applicable to suddenly switched-on potentials 
but not to the adiabatically switched-on potentials. \\ 

Like in the case of static DFT, in TDDFT also an interacting many-electron system
is mapped onto a fictitious non-interacting Kohn-Sham system with the same time-
dependent density as the interacting system. In Floquet formulation \cite{dg,barto} 
the Kohn-Sham system is developed in terms of the periodic time-dependent density only. 
On the other hand, in the the RG formalism the Kohn-Sham wavefunction is a functional of 
both the time-dependent density and an initial state \cite{maitra,vanl}. The uniqueness 
of the KS system follows from the one-to-one $\rho(\vec r,t)\to v(\vec r,t)$ mapping. 
However, in a recent work \cite{mb} by Maitra and Burke it has been suggested that a KS 
system within the Floquet formalism is not unique. The non-uniqueness of the KS system 
has been attributed to its functional dependence on the initial state. Further, they 
suggest that the Hohenberg-Kohn like theorem \cite{dg} may not exist in Floquet theory 
of TDDFT because the "ground Floquet state" cannot be defined properly. This appears 
to have created some doubts \cite{au} about the Floquet theory in TDDFT. It is the 
purpose of this paper to analyze the Floquet theory of TDDFT and show that it is perfectly 
valid. The present analysis also brings out subtle differences between the 
RG and Floquet theory and shows the two formulations to be distinct and founded on two 
different principles. As such Floquet theory of TDDFT cannot be thought of as a particular 
case of the RG theory and the RG arguments can't be applied to Floquet formulation. We begin 
with a short description of the foundations of Floquet and RG theory in TDDFT. \\

For a time-dependent (TD) Hamiltonian $\hat H(t)$ periodic in time, {\it i.e.}
\begin{equation}
\hat H(t) = \hat H(t + \tau)~\;,
\label{fl3}
\end{equation}
where the period $\tau$ is positive with $\omega = \frac{2\pi}{\tau}$, 
the TD Scr\"odinger equation (atomic units are used throughout)

\begin{equation}
\left[\hat H(t) - i\frac{\partial}{\partial t} \right] \Psi_n(\vec r,t) = 0~\;,
\label{fl4}
\end{equation}
has solutions \cite{shirley,sambe} of the form
\begin{equation}
\Psi_n(\vec r,t) = \phi_n(\vec r,t)\,e^{-i\varepsilon_n t} ; 
\phi_n(\vec r,t) = \phi_n(\vec r,t + \tau)~\;,
\label{fl5}
\end{equation}
where the time-periodic functions $\phi_n(\vec r,t)$ are termed as the {\it quasienergy} 
eigen states (QES) and $\varepsilon_n$ are referred to as the the {\it quasienergy} eigenvalues. 
The steady-state wavefunctions $\phi_n(\vec r,t)$ are solutions of 
\begin{equation}
\left[\hat H(t) - i\frac{\partial}{\partial t} \right] \phi_n(\vec r,t) = \varepsilon_n
\phi_n(\vec r,t)~\;,
\label{fl6}
\end{equation}
where $\phi_n(\vec r,t)$ are square integrable and $\varepsilon_n$ are real numbers. 
The quasienergies, $\varepsilon_0,\varepsilon_1,..,\varepsilon_i,...$ are defined 
`modulo $(\omega)$'. However, they can be ordered in increasing order $\varepsilon_0,
\varepsilon_1,...$ by making sure that as the strength of the applied time-dependent 
potential goes to zero, each quasi-energy go to its unperturbed counterpart \cite{epstein}. 
Thus as long as $\omega$ is not one of the resonant frequencies, there exist the 
"ground-state energy" $\varepsilon_0$ and "excited-state energies" $\varepsilon_1,...,
\varepsilon_n$ \cite{epstein} in the steady-state formalism (see Appendix A for details). 
The theory of solutions in time-periodic Hamiltonian can be thought of like the stationary-state 
theory but in an extended Hilbert space ($R+T$) that includes in addition to the space dependent 
function, time-periodic functions also. The operator

\begin{equation} 
{\mathcal{H}}(t) = \hat H(t) - i\frac{\partial}{\partial t}~\;,
\label{fl7}
\end{equation}
is called the Hamiltonian for the steady states in the composite Hilbert space, which 
resembles in many way the Hamiltonian for bound-states. The scalar product in this 
space is defined as

\begin{equation}
\left\{\langle \phi|\psi\rangle \right\} = \frac{1}{\tau} \int_0^\tau dt \int
\phi^*(\vec r,t)\psi(\vec r,t) d\vec r
\label{inner}
\end{equation} 
{\it i.e.} in addition to the space integral, an integral over time is also taken. Here the curly 
brackets `$\{~\}$' indicate the time average over a period. Thus the quasienergy functional is 
given by

\begin{equation}
\varepsilon_n[\phi_n] = \left\{\langle\phi_n|{\mathcal{H}}(t)|\phi_n\rangle \right\}~\;.
\label{fl8}
\end{equation}
We point out - and this is important from TDDFT point of view - that the steady-state solutions 
are obtained by an adiabatic switching of the periodic potential. Finally applying the 
variational principle, each state $\phi_n(\vec r,t)$ can also be obtained \cite{sambe} by 
making the expectation value $\left\{\left\langle\phi_n\left|\hat H(t) - i\frac{\partial}
{\partial t}\right|\phi_n\right\rangle \right\}$ stationary. The stationarity principle becomes 
a minimum principle for the "ground-state" (Here the "ground-state" refers to a
steady-state having the lowest quasienergy.) by assuming completeness of the set $\left\{
\phi_n \right\}$ over the ($R+T$) space. The general proof of the minimum principle
for the ground-Floquet state is given in Appendix~A. \\

Using the minimum property of the ground-state quasienergy functional a 
Hohenberg-Kohn like theorem can be proved \cite{dg} for time-dependent periodic
densities also. We reproduce the proof below. Let us consider the nondegenerate  
ground-state of a system characterized by the Hamiltonian $\hat H(t)$. Suppose there 
exist two different Floquet ground-states $\phi$ and $\phi'$ corresponding to the steady state 
Hamiltonians $\mathcal{H}$ and $\mathcal{H}'$, or more specifically corresponding to the 
external potentials $v(\vec r ,t)$ and $v'(\vec r ,t)$, both of which give rise to the same TD 
density $\rho(\vec r ,t)$. Using the minimal property for the ground-state we get

\begin{equation}
\varepsilon' < \varepsilon + \frac{1}{\tau}\int_{t_1}^{t_2} dt \int [v'(\vec r ,t)
- v(\vec r ,t)] \rho(\vec r ,t) d^3r~\;.
\label{fl11}
\end{equation}
Similarly for the ground-state of the Hamiltonian $\mathcal{H}$

\begin{equation}
\varepsilon < \varepsilon' + \frac{1}{\tau}\int_{t_1}^{t_2} dt \int [v(\vec r ,t)
- v'(\vec r ,t)] \rho(\vec r ,t) d^3r~\;.
\label{fl12}
\end{equation}
Adding Eqs. (\ref{fl11}) and (\ref{fl12}) results in the absurdity,
\begin{equation}
\varepsilon + \varepsilon' < \varepsilon' + \varepsilon~\;.
\label{fl13}
\end{equation}
Hence two different potentials $v(\vec r ,t)$ and $v'(\vec r ,t)$ cannot give the same density 
$\rho(\vec r ,t)$, which implies that the time-dependent potential $v(\vec r ,t)$ is a unique 
functional of the time-dependent density $\rho(\vec r ,t)$. Accordingly, $\mathcal{H}$,$H$ and 
$\phi$ and indeed any "ground-state" property are all unique functional of $\rho(\vec r ,t)$ . Thus 
for Floquet states the HK theorem is \cite{dg} : "the density corresponding to the ground-state 
quasienergy of a time-periodic Hamiltonian determines the corresponding external time-periodic 
potential $v_{ext}(\vec r,t)$ uniquely". For example in perturbation theory up to the second order 
in energy, $E^{(2)}$ is minimum \cite{arup} with respect to $\rho^{(1)}$ for frequencies less 
than the first excitation frequency ({\it i.e.} $\omega < \omega_{10} = E_1 - E_0$, where $E_0$ and 
$E_1$ are the unperturbed ground and excited-state energies). Thus in the linear response 
regime, the theorem would apply to $\rho^{(1)}$ up to frequency $\omega < \omega_{10}$. This 
theorem is then the foundation of Floquet theory of TDDFT \cite{chu1,chu2,chu3,chu4,arup,chu5}. 
We point out that the theory does not have any initial state dependence but requires adiabatic 
switching of the applied external potential. Assuming the $v-$representability, an equivalent 
Kohn-Sham system for an $N-$electron system can also be developed. \\

As pointed out earlier, in RG theory \cite{rg} the proof of the density-to-potential mapping
for time-dependent systems is based directly  on the TD Schr\"odinger equation. In this theory
, it is assumed that the time-dependent potential $v(\vec r,t)$ is turned on at a particular
time $t = t_0$ and all systems are taken to evolve from the same initial state wavefunction
$\Psi(t_0)=\Psi_0$. It is then shown that if the potential has a Taylor series expansion
around the initial time $t=t_0$, then the time-dependent density $\rho(\vec r ,t)$ determines 
the potential $v(\vec r,t)$ uniquely. In turn $\Psi(\vec r ,t)$ is also determined by 
$\rho(\vec r ,t)$. Thus the RG theory is quite distinct from the Hohenberg-Kohn theorem or its 
steady-state DFT counterpart (Deb-Ghosh theorem proved above), which are based on the minimization 
of energy (or quasienergy in the Floquet formalism). Notice that because of the Taylor series 
expansion requirement of $v(\vec r,t)$, RG theorem is applicable to suddenly switched-on
potentials but not to adiabatically switched-on potentials. Table~I gives a comparison
of the two theories.\\

It is clear from the comparison given in Table~I that the two theories are quite distinct 
and their domain of application is also different. As such Floquet formulation of TDDFT 
\cite{dg,barto,chu1,chu2,chu3,chu4,chu5,arup} is not a particular case of the RG theory. 
In light of this we now analyze the paper by Maitra and Burke \cite{mb} and then go on to 
comment on what does the distinction between the Floquet and the RG theory mean in the context of 
calculation of quantities like polarizabilities $\alpha(\omega)$ and excitation energies etc. 
using TDDFT. \\

\begin{table}
\caption{Comparison of Floquet theory and RG theory of TDDFT}
\vspace{0.2in}
\begin{tabular}{|l|l|}
\hline
{~~~~~~~~~\bf Floquet Theory} & {~~~~~~~~~\bf RG Theory} \\
\hline
$\bullet$ ~Based on the minimum principle & $\bullet$ ~Based on the evolution of \\  
of the ground-state energy. & wavefunction from $t = t_0$. \\

$\bullet$ ~Steady-state is reached by &  $\bullet$ ~Wavefunctions are obtained \\ 
adiabatic switching of the & from potentials analytic a $t = t_0$. \\
time-periodic potential. &Adiabatic switching is ruled out. \\

$\bullet$ ~$v(\vec r,t)$ is functional of $\rho(\vec r,t)$ only. & $\bullet$ ~$v(\vec r,t)$ is 
functional of $\rho(\vec r,t)$ as\\ & well as $\Psi_0$.\\ 
\hline
\end{tabular}
\end{table}

Maitra and Burke consider the example of two non-interacting electrons in a
one-dimensional harmonic oscillator potential subject to an adiabatically switched-on
time-dependent potential periodic in time. Thus the Hamiltonian they consider is

\begin{equation}
H = -\frac{1}{2}\frac{d^2}{dx^2} + \frac{1}{2}\omega_0^2x^2 + \lambda x Sin(\omega t)~\;.
\label{fl15}
\end{equation}
Due to the lack of many exactly known analytic solutions of time-dependent
systems, we also consider the same system. The QESs for the above mentioned 
Hamiltonian are known analytically \cite{breu3} to be

\begin{equation}
\phi_n(x,t) = \psi_n({\overline x}(t))e^{\left\{i(\omega A x Cos(\omega t) + \alpha(t)\right\}}~\;,
\label{fl16}
\end{equation}
where $\psi_n$ are the eigenstates of the static Harmonic oscillator and ${\overline x}(t)=
x - A Sin(\omega t)$. The parameters of the wavefunction are given as  

\begin{equation}
A = \frac{\lambda}{\omega^2 - \omega_0^2}
\label{fl17}
\end{equation}
and
\begin{eqnarray}
\alpha(t) = \lambda^2\frac{\left\{\frac{Sin(2\omega t)}{8} + Cos(\omega t) - 1 - \omega^2
Cos(\omega t)\times\frac{Sin(\omega t)}{(\omega^2 - \omega_0^2)}\right\}}{\omega(\omega^2 - 
\omega_0^2)} 
\label{fl18}
\end{eqnarray}
The corresponding quasienergy for the QES $\phi_n(x,t)$ is 

\begin{equation}
\varepsilon_n = \left[(n + \frac{1}{2})\omega_0 + \frac{\lambda^2}{4(\omega^2 - \omega_0^2)}
\right]~modulo(\omega)
\label{fl19}
\end{equation}
We first point out that the quasienergies have been expressed in such a manner that as 
$\lambda \to 0$, the energy eigenvalues go to their respective time-independent eigenvalues. 
Thus energy defined `modulo$(\omega)$' does not create any difficulty. Secondly, as long as
$\omega \neq \omega_{0}$, there is clearly a well defined "ground-state" in these
solutions. Thus $\varepsilon_0 = \frac{\omega_0}{2} + \frac{\lambda^2}{4(\omega^2 - 
\omega_0^2)}$ represents the "ground-state" energy of the system, whereas the other 
energies are the "excited-state" energies.

Maitra and Burke consider a singlet state with one electron occupying the ground-state
({\em i.e.} $n = 0$) and the other the first excited-state ({\em i.e.} $n = 1$)
quasienergy orbitals. This gives the density of the system to be

\begin{equation}
\rho(x,t) = \sqrt{\frac{\omega_0}{\pi}}\left(1 + 2\omega_0{\overline x}(t)^2 \right)
e^{-\omega_0{\overline x}(t)^2}
\label{fl20}
\end{equation}
Now they generate the same density by another system with a different periodic potential. 
For doing this they consider a Floquet state which is also a spin singlet but with one 
doubly occupied steady state. Thus

\begin{equation}
\tilde \Phi(x_1,x_2,t) = \tilde \phi(x_1,t) \tilde\phi(x_2,t)e^{2i{\tilde\varepsilon}t}~\;,
\label{fl21}
\end{equation}
with
\begin{equation}
\tilde \phi(x,t) = \sqrt{\frac{\rho(x,t)}{2}}e^{i\beta(x,t)}~\;,
\label{fl22}
\end{equation}
where $\beta(x,t)$ is a real time-periodic function, $\beta(x,t+\tau)=\beta(x,t)$.
By inverting the TD Schr\"odinger equation one will have a different potential

\begin{equation}
\tilde v(x,t) = \frac{1}{2}\frac{\tilde \phi''(x,t)}{\tilde \phi(x,t)} + 
i\frac{\dot{\tilde \phi}(x,t)}{\tilde \phi(x,t)} + {\tilde\varepsilon}
\label{fl23}
\end{equation}
In this $\tilde{\varepsilon}$ is the orbital quasienergy for the second Floquet
state. The function $\beta(x,t)$ is determined by assuming the potential and
quasienergy to be real. Now the new potential and phase for the second
Floquet state are given by

\begin{eqnarray}
\tilde v(x,t) &=& \frac{1}{2}\omega_0^2{\overline x}(t)^2\left\{ 1 - \frac{4}{(1 + 2 \omega_0
{\overline x}(t)^2)^2} - \frac{4}{1 +2 \omega_0{\overline x}(t)^2 }\right\} + \nonumber\\
&&\frac{\omega_0}{1 +2 \omega_0{\overline x}(t)^2} + \omega_0 - \frac{\omega_0^2A^2 Sin^2(\omega t)}{2}
\nonumber\\
\beta(x,t) &=& A\omega Cos(\omega t) x - A^2\left(\omega^2 - \frac{\omega_0^2}{2}\right)
\frac{Sin(2\omega t)}{4\omega} 
\label{fl24}
\end{eqnarray}
The phase $\beta(x,t)$ is uniquely determined up to a purely time-dependent function. At large
$x$ both the potentials $v(\bar x,t),v'(\bar x,t)\to \frac{\omega_0^2x^2}{2} + 
\lambda x Sin(\omega t)$ and the quasienergy is $\tilde \varepsilon = \frac{3\omega_0}
{2} + \frac{A^2\left(\omega^2 - \omega_0^2/2\right)}{2}~~modulo(\omega)$ \\



\begin{figure}[thb]
\includegraphics[width=4.5in,angle=0.0]{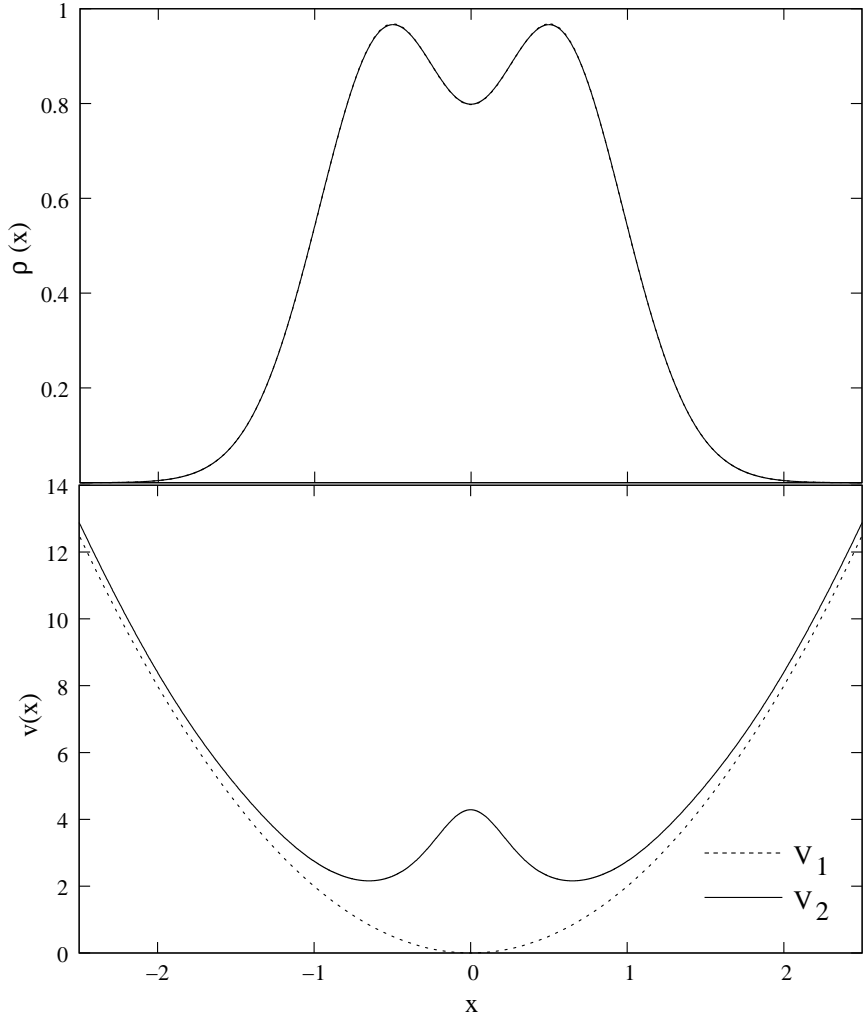}
\caption{Shown in the figure are the two potentials (lower panel) giving rise to 
the same excited-state density (upper panel)of a harmonic 1D oscillator by 
putting one electron in the ground and other electron in the first excited-state.}
\label{combine}
\end{figure}

Based on the fact that they have been able to generate one density from two different
potentials, Maitra and Burke then conclude that this is a manifestation of the initial 
state-dependence of TDDFT. Thus they find Floquet theory of TDDFT to be incomplete. We 
now show that Maitra and Burke are incorrect on two counts. First, they have considered 
a system in steady-state solution that is supposed to have been obtained  by an adiabatic 
switching. Thus RG theory is not applicable to this system. Therefore one cannot talk
about the dependence on the initial state which is a specific feature of the RG
theory. The problem falls in the domain of steady-state solutions and as such 
should be looked at within the Floquet theory of TDDFT. As discussed earlier,
the domain of applicability of the Floquet theory of TDDFT is precisely the
kind of example that has been considered above. Secondly, the Floquet theory of TDDFT
is similar to the stationary-state DFT and as the derivation earlier shows,
it is applicable to only the ground-states of the steady-states. Maitra and Burke
on the other hand have applied it to an excited-state of the steady-state solution and shown
that an excited-state density can be generated by two different potentials.
That however, does not invalidate Floquet theory of TDDFT. This point needs
further elaboration and we do that in the following.\\

\begin{figure}[thb]
\includegraphics[width=4.5in,angle=0.0]{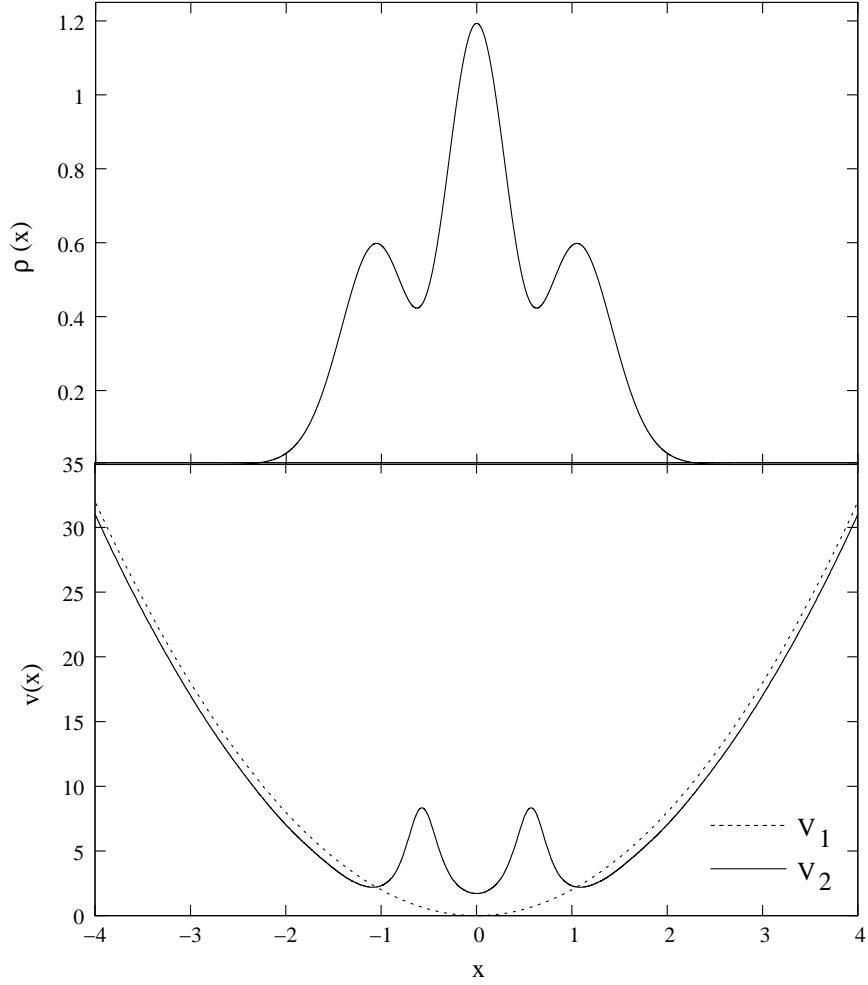}
\caption{Two potentials (lower panel) yielding the same excited-state
density (upper panel) for an excited state of the two-electron in an 1D
harmonic oscillator.}
\label{differ}
\end{figure}

For our discussion, we show that even in the stationary-state theory, we can generate 
a given excited-state density by two different potentials \cite{smss,harb,gb,shh,sh}. Again
consider two electrons in a one-dimensional harmonic oscillator potential. If the
two electrons are in the ground-state, the potential is determined uniquely by the
Hohenberg-Kohn theorem. However, for excited-states, there is no such theorem
and more than one potentials can give the same density. For this let us consider one 
electron in $n = 0$ and one in $n = 1$ state. The corresponding density is shown in 
the upper panel of Fig.~1. One potential corresponding to this density is obviously the 
harmonic potential $v(x) = \frac{1}{2}\omega_0^2 x^2$ itself. Another potential can be 
generated by putting both the electrons in the ground-state orbital and then by inverting 
the Schr\"odinger equation or by using the van-Leeuwen-Baerends method \cite{lb}. These two 
potentials are shown in the lower panel of Fig.~1 . We consider one more example similar to 
that discussed above, where the excited-state density of two non-interacting electrons moving 
in a one-dimensional harmonic oscillator potential is obtained by putting one electron in 
$n = 0$ and the other in $n = 2$ state. Now this density (shown in the upper panel of Fig.~2) 
is produced by an alternative potential, which is shown in the lower panel of Fig.~2 along 
with the original potential. Thus if a system is in excited-state, more than one potentials 
can give the same density. On the other hand for the ground-state, occupation is uniquely 
defined and so is the potential. More examples of this kind exist in the literature 
\cite{smss,harb,gb,shh,sh}. Exactly in the same manner as in stationary-state theory, in Floquet 
theory of TDDFT also the "ground-state" gives the potential uniquely but for "excited-states" 
more than one potential can be found. This is precisely what has been done in \cite{mb}. 
Maitra and Burke do state in their work that their results resemble those of excited-states in 
time-independent problems but fail to make further connection with the latter.\\

Having explained the work of Maitra and Burke, we now also comment on the TDDFT
calculations in light of the Floquet and RG theories of TDDFT. It is clear that in
calculating frequency dependent polarizability $\alpha(\omega)$, it is the steady-state
of a system that is employed. Thus in calculation of $\alpha(\omega)$, and related
quantities such as excitation energies and oscillator strengths, \cite{casida} it is the Floquet
theory of TDDFT rather than the RG theory that is being applied. Finally, one may
raise a question if Floquet theory of TDDFT is applicable only to "ground-states".
The answer is that even for steady-state "excited-states", a theory similar to the
stationary-state excited-state DFT \cite{ag1,ag2,ln,smss,gb,harb,shh,sh} can be developed 
but that is not our main concern here.\\

To conclude, we have shown that Floquet theory of TDDFT is well founded and is
distinct from the RG theory. Further, its recent criticism by Maitra and Burke
is easily explained on the basis of a careful analysis of Floquet theory and
RG theory.\\

{\bf Acknowledgement:} We thank Professors B. M. Deb and K.D. Sen and Drs. S. K. Ghosh and
Arup Banerjee for their comments on the manuscript. Critical remarks of Dr. Telnov are
also acknowledged with pleasure. Discussions with Dr. K. Srihari are also acknowledged.

\appendix
\section{}
In this appendix, we show that if one confines the quasienergies to a range such that they
go to their unperturbed counterpart when the time-periodic potential is turned 
off, then there is a well defined ground-state quasienergy satisfying the 
minimum variational principle. We first point out that defining the eigenenergies
as suggested above is equivalent to taking the corresponding quasienergy state 
such that it has no free time-dependent factor of the form $\left(\sum_p~e^{ip\omega t}
\right)$, where $\left\{p = 0,\pm1,\pm 2, ...., \pm i,..\right\}$, associated with it 
as the strength of the time-periodic potential goes to zero . Let us denote such quasienergy 
states as $\phi_0(\vec r,t),\phi_1(\vec r,t),...,\phi_i(\vec r,t),... $ with the corresponding 
eigenvalues $\varepsilon_0,\varepsilon_1,...,\varepsilon_i,...$ such that $\varepsilon_0 < 
\varepsilon_1 < ... <\varepsilon_i < ...$. The quasienergy states satisfy the normalization 
condition

\begin{equation}
\int \phi_m^\dagger(\vec r,t)\phi_n(\vec r,t) d\vec r = \delta_{mn}~\;,
\label{norm}
\end{equation}
as is easily seen by the hermiticity of the Hamiltonian coupled with its adiabatic
switching.

Let us consider a Hamiltonian $\hat H(\vec r,t) = \hat H_0(\vec r) + \hat v
(\vec r,t)$ with $\hat v(\vec r,t)$ being time periodic. Now a trial periodic function 
$\Phi(\vec r,t)$ can be expanded in terms of unperturbed states $\bar\phi_i(\vec r)$'s 
of $\hat H_0(\vec r)$ and its associated states $\bar\phi_i(\vec r)~e^{ip\omega t}$ as 

\begin{eqnarray}
\Phi(\vec r,t) &=& a_0 \bar\phi_0(\vec r) + a_1 \bar\phi_1(\vec r) + .... +
a_i \bar\phi_i(\vec r) + ..... \nonumber\\
&& + a_0^{(\omega)} \bar\phi_0(\vec r) e^{i\omega t} + a_1^{(\omega)} \bar\phi_1(\vec r) 
e^{i\omega t} + .... + a_i^{(\omega)} \bar\phi_i(\vec r) e^{i\omega t} + ..... \nonumber\\
&& + a_0^{(2\omega)} \bar\phi_0(\vec r) e^{i 2\omega t} + a_1^{(2\omega)} \bar\phi_1(\vec r) 
e^{i 2\omega t}+ .... + a_i^{(2\omega)} \bar\phi_i(\vec r) e^{i 2\omega t} + ..... \nonumber\\
&=& \left( \sum_p a_0^{(p\omega)} e^{i p\omega t}\right)\bar\phi_0(\vec r) +
\left(\sum_p a_1^{(p\omega)} e^{i p\omega t}\right) \bar\phi_1(\vec r) + .....  \nonumber\\ 
&& + \left(\sum_p a_i^{(p\omega)} e^{i p\omega t}\right)\bar\phi_i(\vec r) + ......  
\label{md1}
\end{eqnarray}
In the expansion above, each coefficient $ a_i^{(p\omega)}$~$(p \neq 0)$ should become
zero as $\hat v(\vec r,t) \to 0$. Otherwise coefficient $ a_i^{(p\omega)}$ would have 
two components: one arising from the applied time-dependent potential causing unperturbed states 
to mix and the other is the coefficient of physically equivalent state $\phi_i e^{ip\omega t}$.
To see this let us add $\Phi(\vec r,t)e^{i\omega t}$ to $\Phi(\vec r,t)$ to get 

\begin{eqnarray}
\Phi'(\vec r,t) &=& a_0 \bar\phi_0(\vec r) + a_1 \bar\phi_1(\vec r) + .... +
a_i \bar\phi_i(\vec r) + ..... \nonumber\\
&& + \left(a_0^{(\omega)} + a_0\right) \bar\phi_0(\vec r) e^{i\omega t} + \left(a_1^{(\omega)} 
+ a_0\right)\bar\phi_1(\vec r)e^{i\omega t} + .... + \left(a_i^{(\omega)} +a_0\right) 
\bar\phi_i(\vec r) e^{i\omega t} + ..... \nonumber\\ 
&=& \left( 1 + e^{i\omega t}\right)\Phi(\vec r,t)
\label{norm1}
\end{eqnarray}
For normalized wavefunctions the factor in front of $\Phi(\vec r,t)$ gives rise to a purely 
time-dependent phase factor and can therefore be ignored. This is best illustrated if we 
look at the unperturbed time-independent problem in the Floquet formulation.

In the case of time-independent problem in Floquet formulation the time-dependent
wavefunction corresponding to a state $\phi_i(\vec r)$ can be written in two equivalent
forms as

\begin{equation}
\Psi_i(\vec r,t) = e^{-iE_it}\phi_i(\vec r) \equiv e^{-i(E + p\omega)t} \underbrace{
\phi_i(\vec r)~e^{+ip\omega t} = \phi_i(\vec r,t)}~\;,
\label{mod1}
\end{equation}
where $\phi_i(\vec r,t)$ are the Floquet states satisfies

\begin{equation}
\left( \hat H(t) - i \frac{\partial}{\partial t}\right)\phi_i(\vec r,t) = (E_i + p\omega)
\phi_i(\vec r,t)~\;.
\label{mod2}
\end{equation}
However, each $\phi_i(\vec r,t)$ gives the same $\Psi_i(\vec r,t)$. So they are all
physically equivalent. If we were to expand a trial Floquet function in terms
of $\left\{\phi_i(\vec r)\right\},\left\{\phi_i(\vec r)e^{i\omega t}\right\},....,
\left\{\phi_i(\vec r)e^{i p\omega t}\right\}$, all it will do is give a complicated 
time-dependent phase factor in front of $\phi_i(\vec r)$. How does this physical 
equivalence gets reflected variationally is discussed next.

Let us take a trial periodic wavefunction as \cite{shirley} for the unperturbed problem
as 

\begin{equation}
\Phi(\vec r,t) = \sum_{ip} a_{i}^{(p)}\phi_i(\vec r)~e^{i p\omega t}~\;,
\label{mod3}
\end{equation}
with all $\phi$'s and their equivalent states included. The quasienergy of the system 
is

\begin{eqnarray}
&&\frac{1}{T} \int \left\langle \Phi(\vec r,t)\left| \hat H(t) - i \frac{\partial}
{\partial t}\right|\Phi(\vec r,t)\right\rangle dt \nonumber\\
&=& \frac{1}{T} \int \sum_{ip,jq} a_{i}^{(p)\star}\phi_i^\star(\vec r,t)~e^{-i p\omega t}
(E_j + q\omega) a_{j}^{(q)}\phi_j(\vec r,t)~e^{i q\omega t} d\vec r dt \nonumber\\
&=& \sum_{ip,jq} \delta_{ij} \delta_{pq} (E_j + q\omega) a_{i}^{(p)\star}a_{j}^{(q)}\nonumber\\
&=& \sum_{ip} \left|a_{i}^{(p)}\right|^2 (E_i + p\omega)~\;.
\label{mod4}
\end{eqnarray}
Since $\Phi(\vec r,t)$ is normalized at all times, to obtain $a_i^{(p)}$, the functional
above is made stationary \cite{arup} with the condition 

\begin{equation}
\frac{1}{T}\int \left\langle \Phi(\vec r,t)|\Phi(\vec r,t)\right\rangle =
\sum_{ip} \left|a_{i}^{(p)}\right|^2 = 1~\;.
\label{mod5}
\end{equation}
Using the techniques of Lagrange multipliers this leads to the equation

\begin{equation}
a_i^{(p)}\left[E_i + p\omega - \mu \right] = 0~\;,
\label{mod6}
\end{equation}
where $\mu$ is the Lagrange multiplier and gives the quasienergy of the system. The
equation above gives $\mu = E_i + p\omega$ with $a_i^{(p)} \neq 0$ but all other
$a_i^{(p)}$'s vanishes. Thus stationary variational procedure picks Floquet states
from only one particular zone ({\em i.e.} '$p$' is fixed) to represent the system.
However, irrespective of which '$p$' we take, the final wavefunction is the same. 
Thus Floquet states from different zones do not represent different states. As such 
one may restrict oneself to only one zone in the expansion of the Floquet state.
Question that arise is: does restricting oneself to one particular zone lead
to a minimum principle? Actually, it does as we will now show. It is best shown
for $p = 0$ zone but the result is true for any '$p$'. Thus
   
\begin{equation}
\Phi(\vec r,t) = \sum_{i,p=0} a_{i}^{(0)}\phi_i(\vec r)~\;,
\label{mod7}
\end{equation}

\begin{equation}
\langle \Phi|\hat H - i\frac{\partial}{\partial t}|\Phi\rangle = \sum_i \left|
a_{i}^{(0)}\right|^2 \varepsilon_i ~>~ \sum_i \left|a_{i}^{(0)}\right|^2
\varepsilon_0~\;,
\label{mod8}
\end{equation}
by the standard variational argument. 

Having discussed the time-independent case, we now discuss the time-dependent case.
In this case the basis functions are the Floquet state $\phi_i(\vec r,t)$ of the 
full Hamiltonian and their physically equivalent counterparts $\phi_i(\vec r,t) 
e^{ip\omega t}$. If we expand a trial Floquet state $\Phi(\vec r,t)$ as

\begin{equation}
\Phi(\vec r,t) = \sum_{ip} c_{i}^{(p)}\phi_i(\vec r,t)~e^{i p\omega t}~\;,
\label{mod9}
\end{equation}
we again argue that including Floquet states from all zones does not really give
us any new information. So the expansion should be restricted to only one zone.
Mathematically it is again shown as was done above. The approximate quasienergy
is given as 

\begin{equation}
\frac{1}{T}\int \langle \Phi|\hat H - i\frac{\partial}{\partial t}|\Phi\rangle dt = 
\sum_{i,p^\prime} \left|c_i^{(p)} \right|^2 (\varepsilon_i + p \omega)~\;.
\label{mod10}
\end{equation}

So the stationarity \cite{arup} of the time averaged expectation value in Eq.\ref{mod10} 
under the constraint $\frac{1}{T}\int\left\langle\Phi|\Phi \right\rangle dt = 1$ gives

\begin{equation}
c_i^{(p)}\left\{\varepsilon_i + p\omega - E \right\} = 0
\label{mod11}
\end{equation}
implying again that $E = \varepsilon_i + p\omega$ with $c_i^{(p)}$ - only in one zone.
However, Floquet states from each zone give the same wavefunction $\Psi = e^{-i(E +
p\omega)t} \phi_i(\vec r,t) e^{ip\omega t}$. Thus in expanding an approximate
$\Phi(\vec r,t)$ one can restrict oneself to one particular zone. This leads to a 
minimum principle following the standard arguments.

The question arises how do we make sure that the trial wavefunction comprises
Floquet states from one particular zone only. This is best done for $p = 0$
zone by making sure that if we take $v(\vec r,t) = 0$, the trial wavefunction
must be time-independent. If Floquet states from other zones are also present
in the trial wavefunction, it will not become time-independent as the 
time-dependent potential becomes zero. With such a trial wavefunction the 
quasienergy follow a minimum principle as shown above.


\end{document}